\documentclass[aps,prd,nofootinbib,amsmath,amssymb,twocolumn,superscriptaddress,10pt]{revtex4}

\pdfoutput=1

\usepackage{graphicx}
\usepackage{dcolumn}
\usepackage{bm}
\usepackage{amssymb}
\usepackage{latexsym}
\usepackage{booktabs}
\usepackage{amsmath}
\usepackage{multirow}
\usepackage[colorlinks=true, linkcolor=red, citecolor=blue]{hyperref}

\newcommand{\be}{\begin{equation}}
\newcommand{\ee}{\end{equation}}
\newcommand{\bq}{\begin{eqnarray}}
\newcommand{\eq}{\end{eqnarray}}

\usepackage[usenames,dvipsnames]{xcolor}

\DeclareMathAlphabet\mathbfcal{OMS}{cmsy}{b}{n}
\definecolor{darkgreen}{cmyk}{0.85,0.2,1.00,0.2}
\definecolor{purple}{cmyk}{0.5,1.0,0,0}

\def\barray{\begin{array}}
\def\earray{\end{array}}
\def\be{\begin{equation}}
\def\ee{\end{equation}}
\def\ben{\begin{equation} \nonumber}
\def\een{\end{equation}}
\def\ban{\begin{eqnarray*}}
\def\ean{\end{eqnarray*}}
\def\ba{\begin{eqnarray}}
\def\ea{\end{eqnarray}}

\def\({\left(}
\def\){\right)}

\begin{document}

\title{Search for sterile neutrinos in holographic dark energy cosmology: Reconciling Planck observation with the local measurement of the Hubble constant}
\author{Ming-Ming Zhao}
\affiliation{Department of Physics, College of Sciences,
Northeastern University, Shenyang 110004, China}
\author{Dong-Ze He}
\affiliation{Department of Physics, College of Sciences,
Northeastern University, Shenyang 110004, China}
\author{Jing-Fei Zhang}
\affiliation{Department of Physics, College of Sciences,
Northeastern University, Shenyang 110004, China}
\author{Xin Zhang\footnote{Corresponding author}}
\email{zhangxin@mail.neu.edu.cn}
\affiliation{Department of Physics, College of Sciences,
Northeastern University, Shenyang 110004, China}
\affiliation{Center for High Energy Physics, Peking University, Beijing 100080, China}

\begin{abstract}

We search for sterile neutrinos in the holographic dark energy cosmology by using the latest observational data. To perform the analysis, we employ the current cosmological observations, including the cosmic microwave background temperature power spectrum data from the Planck mission, the baryon acoustic oscillation measurements, the type Ia supernova data, the redshift space distortion measurements, the shear data of weak lensing observation, the Planck lensing measurement, and the latest direct measurement of $H_0$ as well. We show that, compared to the $\Lambda$CDM cosmology, the holographic dark energy cosmology with sterile neutrinos can relieve the tension between the Planck observation and the direct measurement of $H_0$ much better. Once we include the $H_0$ measurement in the global fit, we find that the hint of the existence of sterile neutrinos in the holographic dark energy cosmology can be given. Under the constraint of the all-data combination, we obtain $N_{\rm eff}= 3.76\pm0.26$ and $m_{\nu,\rm sterile}^{\rm eff}< 0.215\,\rm eV$, indicating that the detection of $\Delta N_{\rm eff}>0$ in the holographic dark energy cosmology is at the $2.75\sigma$ level and the massless or very light sterile neutrino is favored by the current observations.

\end{abstract}
 \maketitle

\section{Introduction}\label{sec:intro}

Recently, Riess et al.~\cite{Riess:2016jrr} reported their new result of improved determination of the Hubble constant with only $2.4\%$ uncertainty, $H_{0}=73.00\pm1.75 ~\rm km  ~s^{-1} ~Mpc^{-1}$, at the $68\%$ confidence level. The improvement in uncertainty can be attributed to not only the enlarged number of supernova (SN) hosts and Cepheids in the Large Magellanic Cloud (LMC) and Milky Way (MW), but also the reduced systematic deviation in the distance to the NGC 4258 and the LMC. Under such a great improvement, the result of $ H_{0}=73.00\pm1.75 ~\rm km  ~s^{-1} ~Mpc^{-1}$ (R16, hereafter) is considered by far to be the best one among the local determinations of the Hubble constant.

Notwithstanding, there are also some widely acknowledged global determinations of $H_{0}$ derived from the constraint of the cosmic microwave background (CMB) data under the assumption of a baseline $\Lambda$CDM cosmology. For example, we have $H_{0}=69.7\pm2.1 ~\rm km  ~s^{-1} ~Mpc^{-1}$ from WMAP9 \cite{Ade:2015xua}; $H_{0}=68.0\pm0.7 ~\rm km~s^{-1} ~Mpc^{-1}$ from WMAP+BAO \cite{Ade:2015xua}; $H_{0}=69.3\pm0.7 ~\rm km  ~s^{-1} ~Mpc^{-1}$ from WMAP9+ACT+SPT+BAO \cite{Bennett:2014tka}; $H_{0}=67.3\pm1.0 ~\rm km  ~s^{-1} ~Mpc^{-1}$ from Planck TT+lowP \cite{Ade:2015xua}; and $H_{0}=67.6\pm0.6 ~\rm km  ~s^{-1} ~Mpc^{-1}$ from Planck TT+lowP+BAO \cite{Ade:2015xua}. However, when comparing these cosmological estimates with R16, we find that tension between them still exists and is becoming more and more significant.

One possible interpretation for the tension is that some sources of systematic errors in astrophysical measurements are not completely understood. Alternatively, however, there is also an explanation that the base $\Lambda$CDM model is incorrect or should be extended. Indeed, it is possible to alleviate the tensions among many astrophysical observations by invoking new physics. For instance, it has been demonstrated that the Planck tensions with the Hubble constant measurement, the count of rich clusters, and the cosmic shear measurements might hint the existence of sterile neutrinos \cite{Viel:2005qj,Lesgourgues:2006nd,Viel:2006kd,Riess:2014tka,Lesgourgues:2014zoa,Riess:2016jrr,Ade:2015xua,Palazzo:2013me,Ko:2014bka,Archidiacono:2014apa,Zhang:2014nta,Archidiacono:2014nda,Li:2014dja,An:2014bik,Zhang:2014lfa,Zhang:2014ifa,Li:2015poa,Wyman:2013lza,Hamann:2013iba} and that dark energy may not be the cosmological constant $\Lambda$ \cite{Li:2013dha}. Meanwhile, if a sterile neutrino species is added, it is able to enhance the early-time Hubble expansion rate and thus change the acoustic scale precisely observed by Planck, leading the cosmological fit results for $H_{0}$ to be in better agreement with its direct measurement \cite{Ade:2015xua,Ade:2013zuv}. Now that the sterile neutrinos may be capable of affecting the constraints on $H_{0}$, in this study, we will take the sterile neutrino into account, showing how the sterile neutrino helps resolve the tension between R16 and the global determinations of $H_{0}$.

On the other hand, it was realized that the properties of dark energy could as well influence the cosmological constraints on the Hubble constant and thus help to reconcile the tension between the local and global determinations of $H_{0}$ in the literature. For example, Di Valentino et al. \cite{DiValentino:2016hlg} considered a 12-parameter extended cosmological model which allows for a variation of equation of state (EoS) of dark energy and they showed that the tension between R16 and the combination of Planck 2015 data and BAO data can be relieved to some extent; see also \cite{DiValentino:2015wba}. Next, Huang et al. \cite{Qing-Guo:2016ykt} tried to reconcile the tension between Planck 2015 and R16 by modeling dark energy in various manners. In these analyses, they all point to a line that a feasible phantom-like dark energy with an effective EoS can alleviate the current tension. Meanwhile, according to \cite{Zhang:2015uhk}, the holographic dark energy (HDE) model has a phantom behavior, and thus favors a high value of $H_{0}$.

Actually, the HDE model is constructed from the effective quantum field theory combined with the requirement of the holographic principle of quantum gravity, which is expected to provide useful clues for a bottom-up exploration of the quantum theory of gravity, thus attracting widespread theoretical interest. In this consideration, the dark energy density is defined as $\rho_{\rm de}=3c^{2}M_{\rm Pl}^{2}R_{\rm EH}^{-2}$, where $M_{\rm Pl}$ is the reduced Planck mass and $R_{\rm EH}$ is the event horizon size of the universe. Note that $c$ is a dimensionless parameter that determines the evolution of dark energy; see Eqs. $(2.4)$$-$$(2.7)$ in \cite{Zhang:2015rha}. Detailed investigations about the HDE model can be found in \cite{Li:2004rb,Huang:2004ai,Zhang:2005hs,Zhang:2007sh,Zhang:2006av,Zhang:2006qu,Li:2009bn,Xu:2016grp,hde1,hde2,hde3,hde6,hde7,hde8,hde9,hde10,hde11,hde12,hde15,hde16,hde17,hde18,hde19,hde22,Zhang:2014ija,Zhang:2007uh,Zhang:2009un,Li:2009zs,Feng:2016djj}.

The EoS of HDE is expressed as
\begin{equation}\label{eos}
  w=-\frac{1}{3}-{\frac{2}{3}}\frac{\sqrt{\Omega_{\rm de}}}{c}.
\end{equation}
According to this equation, one can easily find that in the early times $w\rightarrow -1/3$ (since $\Omega_{\rm de}\rightarrow0$) and in the far future $w \rightarrow $ $-1/3-2/(3c)$ (since $\Omega_{\rm de}\rightarrow 1$). This explains why the phantom divide ($w=-1$) crossing happens when $c<1$. In fact, the current observations favor $c<1$ at the $7\sigma$ level and an obvious phantom behavior of dark energy with $w\sim-1.1$ today \cite{Zhang:2014ija}. Following the analyses in \cite{Riess:2016jrr,Ade:2015xua,DiValentino:2016hlg}, a dark energy with $w\sim -1.1$ could bring the Planck constraint into better agreement with higher values of the Hubble constant. Therefore, the HDE model is worthy to be seriously considered to reconcile the current tension.

In this paper, we will try to reconcile the tension between the local $2.4\%$ determination of the Hubble constant and its global determination through considering the effects of sterile neutrinos in the holographic dark energy cosmology. This is just the primary aim of the current work. Note that this is the first time to consider sterile neutrinos in the holographic dark energy model. The HDE model with sterile neutrinos considered in this paper is called the HDE+$N_{\rm eff}$+$m_{\nu,\rm sterile}^{\rm eff}$ model, where $N_{\rm eff}$ is the effective number of relativistic species, $m_{\nu,\rm sterile}^{\rm eff}$ is the effective mass of the sterile neutrino, and there is a relation of $m_{\nu, \rm sterile}^{\rm eff}=(\Delta N_{\rm eff})^{3/4}m_{\rm sterile}^{\rm thermal}$ between them. Here, $\Delta N_{\rm eff}=N_{\rm eff}-N_{\rm eff}^{\rm SM}$ (with the standard cosmological prediction $N_{\rm eff}^{\rm SM}=3.046$ \cite{Mangano:2005cc,deSalas:2016ztq}) and $m_{\nu,\rm sterile}^{\rm thermal}$ is the true mass of the sterile neutrino \cite{Ade:2015xua}.

In the cosmological global fits, since the addition of dynamical dark energy will increase the degeneracies in the cosmological parameters, other than the CMB power spectrum, we need to combine some other geometric observations such as the baryon acoustic oscillations (BAO) data, the type Ia supernova (SN) data and the independent measurement of the Hubble constant ($H_{0}$). Here, BAO, SN and $H_{0}$ can help break the degeneracies at the low redshifts, providing strong exploration to the equation of state of dark energy at $z\lesssim1$, and constrain the sterile neutrino mass well. We will also use the large-scale structure observations, including the weak lensing (WL), redshift space distortions (RSD), and CMB lensing data. For related works, see, e.g., \cite{Zhang:2014dxk,Dvorkin:2014lea,Zhang:2014nta,Li:2014dja,Archidiacono:2014apa,Bergstrom:2014fqa,Leistedt:2014sia,Beutler:2014yhv,Mantz:2014paa,DiValentino:2015wba,Rossi:2014nea,DiValentino:2015sam}.

The rest of the paper is organized as follows. In Sec.~\ref{sec:cosmol}, we will describe the methodology and observational data sets we use in this paper. In Sec.~\ref{sec:resul}, we will present the results of the cosmological constraint on the HDE+$N_{\rm eff}$+$m_{\nu,\rm sterile}^{\rm eff}$ model. In Sec.~\ref{discussion}, we will make some discussions in depth for the fitting results. Finally, the conclusion will be given in Sec.~\ref{sec:conclu}.

\section{Methodology and data}\label{sec:cosmol}

In this paper, we place constraints on the HDE cosmology with sterile neutrinos by using the latest observational data.

The base parameter set for the basic seven-parameter HDE model is
\begin{equation}
  \{\Omega_{b}h^{2},~\Omega_{c}h^{2},~100\theta_{\rm MC},~\tau,~c,~n_{s},~\ln[10^{10}A_{s}]\},
\end{equation}
where $\Omega_{\rm b}h^{2}$ and $\Omega_{\rm c}h^{2}$ are the present-day baryon and cold dark matter densities, respectively, $100\theta_{\rm MC}$ is 100 times the ratio between the sound horizon and the angular diameter distance at the time of last-scattering, $\tau$ is the Thomson scattering optical depth due to the reionization, $c$ is the specific model parameter in the HDE model determining the evolution of holographic dark energy, and $n_{s}$ and $A_{s}$ are the spectral index and power amplitude of the primordial curvature perturbations, respectively.

When the massive sterile neutrinos are considered in the model, the parameters $m_{\nu,\rm sterile}^{\rm eff}$ and $N_{\rm eff}$ should also be involved in the calculation. To infer the posterior probability distributions of parameters, we use the latest version of the Monte Carlo Markov Chain pakage \texttt{CosmoMC} \cite{Lewis:2002ah} to do the calculations. Besides, the perturbations in dark energy are also considered in our calculations, and thus the ``parameterized post-Friedmann'' (PPF) framework is employed \cite{ppf1,ppf2,ppf3} (for a generalized version of PPF, see \cite{ppf4,ppf5,ppf6,Guo:2017hea,Zhang:2017ize}).

The observational data sets we use in this work are comprised of CMB, BAO, SN, RSD, WL, and CMB lensing.

\emph{The CMB data}: We use the latest CMB TT angular power spectrum data (TT+lowP) from the 2015 release of Planck \cite{Aghanim:2015xee}.

\emph{The BAO data}: We use the BAO measurements from the 6dFGS ($z=0.1$) \cite{Beutler:2011hx}, SDSS-MGS ($z=0.15$) \cite{Ross:2014qpa}, LOWZ ($z=0.32$) and CMASS ($z=0.57$) DR12 samples of BOSS \cite{Alam:2016hwk}. This combination of BAO data has been used widely and proven to be in good agreement with the Planck CMB data.

\emph{The SN data}: For the type Ia supernova observation, we employ the ``JLA" sample, compiled from the SNLS, SDSS and the samples of several low-redshift SN data~\cite{Betoule:2014frx}.

\emph{The $H_{0}$ data}: We employ the result of Riess et al. \cite{Riess:2016jrr}, which is confirmed and improved from their former determination, with the measurement value $\rm H_{0}=73.00\pm1.75 ~\rm km  ~s^{-1} ~Mpc^{-1}$.

\emph{The RSD data}: We employ the two latest RSD measurements, from which we get the CMASS sample with an effective redshift of $z=0.57$ and the LOWZ sample with an effective redshift of $z=0.32$ \cite{Gil-Marin:2016wya}, respectively. The usage of the RSD data is the same as the prescription given by the Planck Collaboration \cite{Ade:2015xua}.

\emph{The lensing data}: Firstly, we use the cosmic shear measurement of weak lensing from the CFHTLenS survey, which perform tomographic analysis with cosmological cuts, specifically removing the angular scales $\theta<3'$ for the two lowest bin combinations, angular scales $\theta<30'$ for $\rm\xi^{-}$ for the four lowest bins, and $\theta<16'$ for the two highest bins for $\rm\xi^{+}$  \cite{Heymans:2013fya}. We denote the shear measurement as ``WL''. Next, we also use the CMB lensing power spectrum from the Planck lensing measurement. The CMB lensing reconstruction data directly probe the lensing power, thereby also sensitive to the sterile neutrino mass \cite{Ade:2015zua}. We denote the CMB lensing measurement as ``lensing'' in this paper.

\section{Results}\label{sec:resul}

\begin{figure}
\begin{center}
\includegraphics[scale=1.5, angle=0]{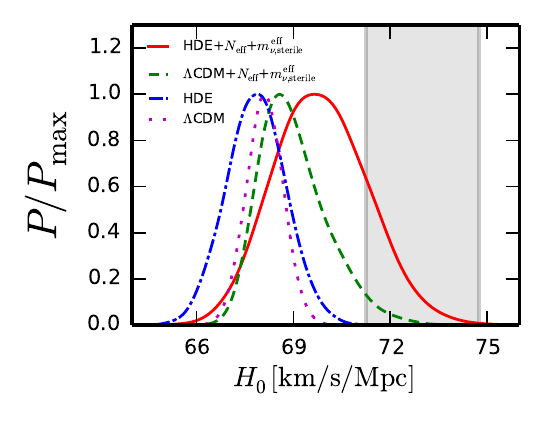}
\caption{The one-dimensional marginalized distributions of $H_0$ for HDE+$N_{\rm eff}$+$m_{\nu,\rm sterile}^{\rm eff}$ (red solid curve), $\Lambda$CDM+$N_{\rm eff}$+$m_{\nu,\rm sterile}^{\rm eff}$ (green dashed), HDE (blue dashed-dotted) and $\Lambda$CDM (purple dotted) under the constraints of the CMB+BAO+SN+RSD+WL+lensing data combination. The result of the latest local measurement of Hubble constant ($H_{0}=73.00\pm1.75 ~\rm km  ~s^{-1} ~Mpc^{-1}$) is shown by the grey band. }
\label{fig:H0}
\end{center}
\end{figure}

\begin{table*}
\caption{\label{tab2}Fitting results for the HDE+$N_{\rm eff}$+$m^{\rm eff}_{\rm \nu,sterile}$, HDE, $\Lambda$CDM+$N_{\rm eff}$+$m^{\rm eff}_{\rm \nu,sterile}$, and $\Lambda$CDM models from the constraint of the data combination CMB+BAO+SN+RSD+WL+lensing. Note that the mass of sterile neutrino $m^{\rm eff}_{\rm \nu,sterile}$ is in units of eV and the Hubble constant $H_{0}$ is in units of $\rm km  ~s^{-1} ~Mpc^{-1}$. The mean values with $\pm1\sigma$ errors are presented, but for the parameters $N_{\rm eff}$ and $m^{\rm eff}_{\rm \nu,sterile}$, the $95\%$ upper limits are given.}
\centering
\begin{tabular}{cccccc}
\hline\hline

 &HDE+$N_{\rm eff}$+$m_{\nu,\rm sterile}^{\rm eff}$&HDE & $\Lambda$CDM+$N_{\rm eff}$+$m_{\nu,\rm sterile}^{\rm eff}$&$\Lambda$CDM\\

\hline

$\Omega_{\rm b}h^2$&$0.02287^{+0.00029}_{-0.00033}$&$0.02249\pm0.00021$&$0.02254^{+0.00024}_{-0.00025}$&$0.02230\pm0.00020$\\
$\Omega_{\rm c}h^2$&$0.1210\pm0.004$&$0.1150^{+0.0015}_{-0.0014}$&$0.1195^{+0.0035}_{-0.0033}$&$0.1178\pm0.0012$\\
$100\theta_{\rm MC}$&$1.04078\pm0.00054$&$1.04137^{+0.00043}_{-0.00042}$&$1.04083^{+0.00053}_{-0.00049}$&$1.04103\pm0.00041$\\
$\tau$&$0.114\pm0.019$&$0.097\pm0.016$&$0.081^{+0.017}_{-0.019}$&$0.063\pm0.013$\\
$n_s$&$0.9788^{+0.0073}_{-0.0111}$&$0.9960^{+0.0120}_{-0.0140}$&$0.9767^{+0.0050}_{-0.0051}$&$0.9688\pm0.0045$\\
${\rm{ln}}(10^{10}A_s)$&$3.166^{+0.040}_{-0.041}$&$3.117\pm0.029$&$3.098^{+0.033}_{-0.041}$&$3.056\pm0.023$\\
$c$&$0.719^{+0.050}_{-0.061}$&$0.682^{+0.035}_{-0.039}$&...&...\\
\hline
$m_{\nu,{\rm{sterile}}}^{\rm{eff}}$&$<0.255$&...&$<0.496$&...\\
$N_{\rm eff}$&$<4.03$&...&$<3.67$&...\\
\hline
$\Omega_{\rm m}$&$0.2981^{+0.0087}_{-0.0086}$&$0.3003^{+0.0085}_{-0.0084}$&$0.3049^{+0.0078}_{-0.0077}$&$0.3035^{+0.0073}_{-0.0072}$\\
$H_0$&$69.80^{+1.40}_{-1.60}$&$67.85\pm0.90$&$69.02^{+0.73}_{-1.28}$&$68.11\pm0.56$\\
$\sigma_8$&$0.805^{+0.017}_{-0.014}$&$0.800\pm0.011$&$0.790^{+0.021}_{-0.017}$&$0.809\pm0.009$\\
\hline\hline
\end{tabular}
\end{table*}

\begin{figure*}
\begin{center}
\includegraphics[scale=1.55, angle=0]{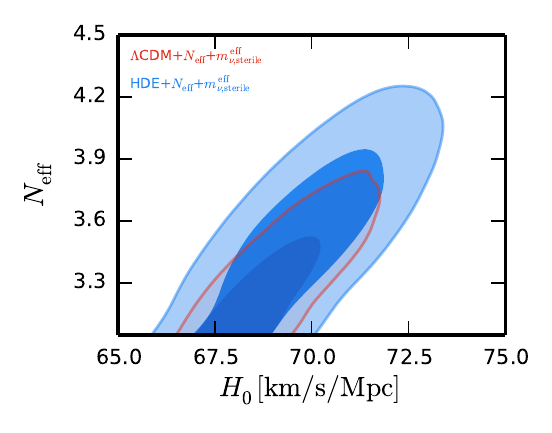}
\includegraphics[scale=1.55, angle=0]{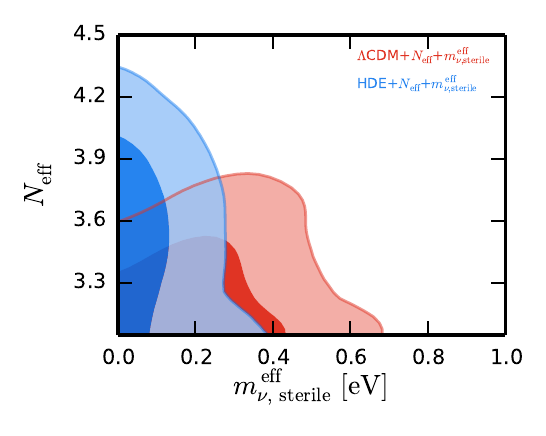}
\caption{Two-dimensional joint, marginalized constraints (68\% and 95\% confidence level) on the $\Lambda$CDM+$N_{\rm eff}$+$m_{\nu,\rm sterile}^{\rm eff}$ (red) and the HDE+$N_{\rm eff}$+$m_{\nu,\rm sterile}^{\rm eff}$ (blue) models from the data combination of CMB+BAO+SN+RSD+WL+lensing. The constraint results in the $H_{0}-N_{\rm eff}$ ({\it left}) and $m _{\nu, \rm sterile}^{\rm eff}-N_{\rm eff}$ ({\it right}) planes are shown.}
\label{fig:HWDE}
\end{center}
\end{figure*}

\begin{figure}
\begin{center}
\includegraphics[scale=1.6, angle=0]{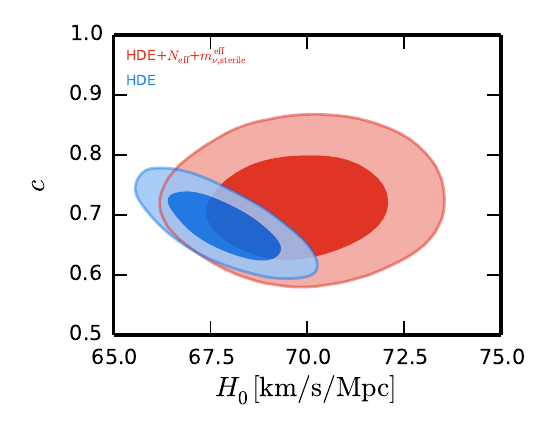}
\caption{Two-dimensional joint, marginalized constraints (68\% and 95\% confidence level) on the HDE+$N_{\rm eff}$+$m_{\nu,\rm sterile}^{\rm eff}$ (red) and HDE (blue) models from the data combination of CMB+BAO+SN+RSD+WL+lensing. The constraint results in the $H_{0}-c$ plane are shown.}
\label{fig:H0c}
\end{center}
\end{figure}

\begin{figure*}
\includegraphics[scale=0.26, angle=0]{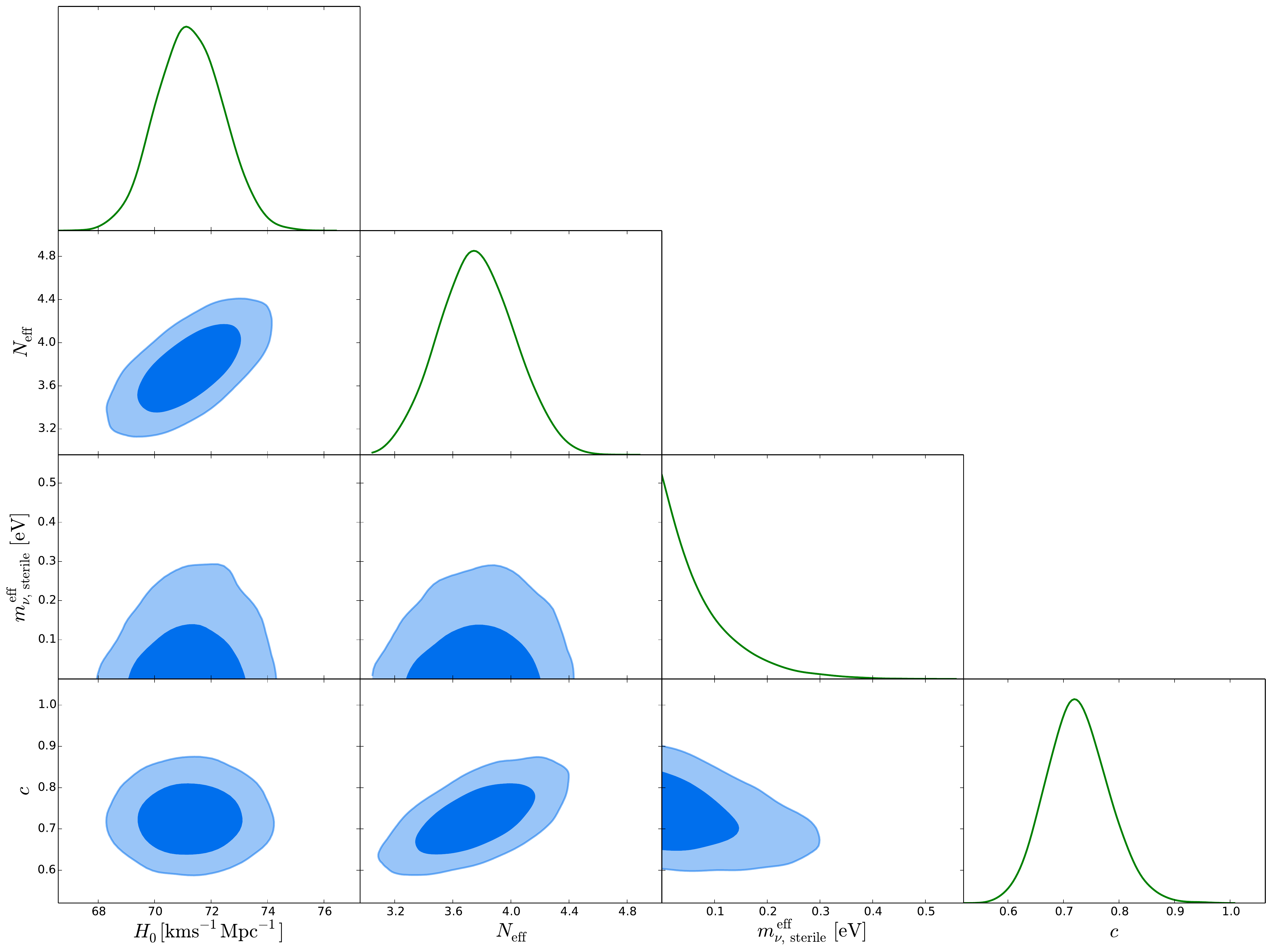}
\caption{Joint, marginalized constraints from CMB+BAO+SN+RSD+WL+lensing+$H_{0}$ on the HDE+$N_{\rm eff}$+$m_{\nu,\rm sterile}^{\rm eff}$ model. One-dimensional marginalized distributions and two-dimensional contours (68\% and 95\% CL) for the parameters $H_{0}$, $N_{\rm eff}$, $m_{\nu,\rm sterile}^{\rm eff}$ and $c$ are shown. }
\label{fig:H0nnuc}
\end{figure*}

\begin{table}
\caption{\label{tab3}Fitting results for the HDE+$N_{\rm eff}$+$m^{\rm eff}_{\rm \nu,sterile}$ model from the constraint of the data combination CMB (Planck TT+lowP)+BAO+SN+RSD+WL+lensing+$H_{0}$. Note that the mass of sterile neutrino $m^{\rm eff}_{\rm \nu,sterile}$ is in units of eV and the Hubble constant $H_{0}$ is in units of $\rm km  ~s^{-1} ~Mpc^{-1}$. }
\centering
\begin{tabular}{ccc}

\hline\hline
$\Omega_{\rm b}h^2$&$0.02304\pm0.00028$\\
$\Omega_{\rm c}h^2$&$0.1238^{+0.0035}_{-0.0036}$\\
$100\theta_{\rm MC}$&$1.04053^{+0.00052}_{-0.00051}$\\
$\tau$&$0.119^{+0.019}_{-0.020}$\\
$n_s$&$1.004\pm0.011$\\
${\rm{ln}}(10^{10}A_s)$&$3.182^{+0.039}_{-0.040}$\\
$c$&$0.728^{+0.052}_{-0.062}$\\
$m_{\nu,{\rm{sterile}}}^{\rm{eff}}$&$<0.215$\\
$N_{\rm eff}$&$3.76\pm0.26$\\
$\Omega_{\rm m}$&$0.2923^{+0.0076}_{-0.0077}$\\
$H_0$&$71.2\pm1.2$\\
$\sigma_8$&$0.814\pm0.013$\\

\hline\hline
\end{tabular}
\end{table}


First of all, we constrain the cosmological parameters by using the data combination of CMB+BAO+SN+RSD+WL+lensing for the HDE+$N_{\rm eff}$+$m^{\rm eff}_{\rm \nu,sterile}$ model, the HDE model, the $\Lambda$CDM+$N_{\rm eff}$+$m^{\rm eff}_{\rm \nu,sterile}$ model, and the $\Lambda$CDM model, respectively. Note that here we do not include the $H_{0}$ data in our data combination because there may exist tension between Planck observation and $H_{0}$ measurement for the $\Lambda$CDM model. In order to make a uniform comparison for these models, we present the detailed fit results in Table~\ref{tab2}. In this table, note here that, we quoted $\pm1\sigma$ errors of the best-fit values for the parameters, but for the parameters (e.g., $N_{\rm eff}$ and $m^{\rm eff}_{\rm \nu,sterile}$) that cannot be well constrained, we give the $95\%$ C.L. upper limits. Moreover, the derived parameters $H_{0}$, $\Omega_{m}$, and $\sigma_{8}$ are also listed here. We show the one-dimensional posterior distributions of $H_{0}$ for the aforementioned four models in Fig.~\ref{fig:H0}. From this figure, we can clearly see that once the sterile neutrinos are considered in the HDE model, then under the constraint of the data combination mentioned above the fit value of $H_{0}$ will become much larger. As also can be seen from Table~\ref{tab1}, we obtain $H_{0}=69.80^{+1.40}_{-1.60}\,\rm km\,s^{-1}\,Mpc^{-1}$ for the HDE+$N_{\rm eff}$+$m_{\nu,\rm sterile}^{\rm eff}$ model, $H_{0}=67.85\pm0.90 \,\rm km\,s^{-1}\,Mpc^{-1}$ for the HDE model, $H_{0}=69.02^{+0.73}_{-1.28}\,\rm km\,s^{-1}\,Mpc^{-1}$ for the $\Lambda$CDM+$N_{\rm eff}$+$m_{\nu,\rm sterile}^{\rm eff}$ model, and $H_{0}=68.11\pm0.56\,\rm km\,s^{-1}\,Mpc^{-1}$ for the $\Lambda$CDM model, which indicates that the tensions with the local determination of Hubble constant ($H_{0}=73.00\pm1.75 ~\rm km  ~s^{-1} ~Mpc^{-1}$) are at the $1.43\sigma$ level, $2.64\sigma$ level, $2.10\sigma$ level, and $2.66\sigma$ level, respectively. We find that, in the HDE+$N_{\rm eff}$+$m_{\nu,\rm sterile}^{\rm eff}$ model, the tension is much smaller than the cases of the other three models, which implies that a larger value of $H_{0}$ can be derived once sterile neutrinos and holographic dark energy are both considered.

Next, with the purpose of visually showing how the dark energy properties lead to a higher value of $H_{0}$, in Fig.~\ref{fig:HWDE}, we present the two-dimensional posterior distribution contours in the $H_{0}-N_{\rm eff}$ plane for the models of $\Lambda$CDM+$N_{\rm eff}$+$m_{\nu,\rm sterile}^{\rm eff}$ and HDE+$N_{\rm eff}$+$m_{\nu,\rm sterile}^{\rm eff}$, presented in red and blue, respectively. In the $H_{0}-N_{\rm eff}$ plane, we can clearly see that $N_{\rm eff}$ is in positive correlation with $H_{0}$ for both models. However, through the comparison of these two models, we find that the HDE model can accommodate a larger $N_{\rm eff}$ and a larger $H_0$, compared to the $\Lambda$CDM model (see also Table~\ref{tab2}). Evidently, variations in the dark energy density of HDE cosmology due to its dynamical property could have some effects on the parameters such as $H_{0}$ and $N_{\rm eff}$ because they would change to compensate the angular size observed by Planck. Therefore, the impacts of dark energy lead to the changes in the fitted values of $H_{0}$ and $N_{\rm eff}$.

In addition, dynamical dark energy properties can also exert a significant impact on the parameters of sterile neutrinos, with the constraint result for sterile neutrinos in the $m_{\nu,\rm sterile}^{\rm eff}-N_{\rm eff}$ plane shown in the right panel of Fig.~\ref{fig:HWDE}. We obtain $N_{\rm eff}<3.67$ and $m_{\nu,\rm sterile}^{\rm eff}<0.496\,\rm eV$ for the $\Lambda$CDM+$N_{\rm eff}$+$m_{\nu,\rm sterile}^{\rm eff}$ model, and obtain $N_{\rm eff}<4.03$ and $m_{\nu,\rm sterile}^{\rm eff}<0.225\,\rm eV$ for the HDE+$N_{\rm eff}$+$m_{\nu,\rm sterile}^{\rm eff}$ model. Obviously, in the HDE cosmology, a much tighter constraint on the effective mass of the sterile neutrino can be derived. In the previous study \cite{Zhang:2015uhk} (see also \cite{Wang:2016tsz}), it was found that, in the HDE model, we can obtain the most stringent upper limit on the total mass of active neutrinos, $\sum m_{\nu}<0.113\,\rm eV$, with the data combination of Planck TT,TE,EE+BAO+SN+$H_{0}$+lensing. Similarly, with regard to the sterile neutrino, the HDE cosmology also leads to an extremely stringent upper limit on the effective mass of the sterile neutrino. Besides, in the HDE cosmology, a much larger upper limit on $\Delta N_{\rm eff}$ is also derived, compared to the $\Lambda$CDM cosmology. Thus, in the HDE cosmology, we obtain $m_{\nu,\rm sterile}^{\rm eff}<0.225\,\rm eV$ and $\Delta N_{\rm eff}<0.98$ by using the current CMB+BAO+SN+RSD+WL+lensing data, which also leads to a stringent constraint on the true mass of the sterile neutrino $m_{\nu,\rm sterile}^{\rm thermal}$ (according to $m_{\nu, \rm sterile}^{\rm eff}=(\Delta N_{\rm eff})^{3/4}m_{\rm sterile}^{\rm thermal}$).

On the other hand, to directly show how sterile neutrinos affect the constraints on $ H_{0}$ in the HDE cosmology, we plot the two-dimensional posterior distribution contours ($68\%$ and $95\%$ CL) in the $H_{0}-c$ plane for the HDE+$N_{\rm eff}$+$m_{\nu,\rm sterile}^{\rm eff}$ and HDE models in Fig.~\ref{fig:H0c}. We find that, in the HDE model, $H_{0}$ is obviously anti-correlated with $c$, but in the HDE+$N_{\rm eff}$+$m_{\nu,\rm sterile}^{\rm eff}$ model, the range of $c$ is greatly enlarged and the correlation becomes weaker. For the parameter $c$, we obtain $c=0.719^{+0.050}_{-0.061}$ for the HDE+$N_{\rm eff}$+$m_{\nu,\rm sterile}^{\rm eff}$ model and $c=0.682^{+0.035}_{-0.039}$ for the HDE model. By comparison, we can see that a higher $c$ is favored by the consideration of sterile neutrinos in the HDE scenario, which also leads to the fact that considering sterile neutrinos in cosmology can largely amplify the parameter space, for both $H_{0}$ and $c$.

We have shown that, in the HDE cosmology, once the sterile neutrino is considered, then the tension between the Planck observation and the latest Hubble constant measurement can be greatly relieved (the residual tension is only at the 1.43$\sigma$ level). Therefore, we can further combine the $H_0$ measurement of R16 in the global fit to search for the sterile neutrino in the HDE cosmology. Namely, we use the CMB+BAO+SN+WL+RSD+lensing+$H_{0}$ data combination to constrain the HDE+$N_{\rm eff}$+$m_{\nu,\rm sterile}^{\rm eff}$ model. The constraint results are shown in Fig.~\ref{fig:H0nnuc} and Table~\ref{tab3}.

Figure~\ref{fig:H0nnuc} shows the one- and two-dimensional marginalized posterior distributions of the parameters $H_{0}$, $N_{\rm eff}$, $m_{\nu,\rm sterile}^{\rm eff}$ and $c$, for the HDE+$N_{\rm eff}$+$m_{\nu,\rm sterile}^{\rm eff}$ model, for the constraint of the all-data combination (CMB+BAO+SN+WL+RSD+lensing+$H_{0}$). We can see from the $H_{0}-c$ plane of this figure that there exists practically no obvious correlation between $H_{0}$ and $c$ due to the consideration of sterile neutrinos, which is accordant with the constraint results under the data combination without $H_{0}$. Also, in the $N_{\rm eff}-H_{0}$ plane and the $c-N_{\rm eff}$ plane, we see that $N_{\rm eff}$ is positively correlated with $H_{0}$ and $c$ at the same time (the detailed physics about this can be found in \cite{Zhang:2015uhk,Zhao:2016ecj}), and the effective number of relativistic species $N_{\rm eff}$ can be very successfully constrained under the all-data combination, namely, $N_{\rm eff}= 3.76\pm0.26$, which indicates a detection of $\Delta N_{\rm eff}>0$ at the $2.75\sigma$ level.

As for the effective mass of sterile neutrino $m_{\nu,\rm sterile}^{\rm eff}$, we still only obtain a 95\% CL upper limit, $m_{\nu,\rm sterile}^{\rm eff}< 0.215\,\rm eV$, which is more stringent than the limit derived from the data combination without $H_{0}$. Also, in this case, we obtain $H_{0}=71.20\pm1.20 ~\rm km  ~s^{-1} ~Mpc^{-1}$, with the tension with R16 improved from $1.43\sigma$ to $0.85\sigma$.

In \cite{Feng:2017nss}, it was shown that, in the $\Lambda$CDM cosmology, a search for massive sterile neutrinos with the latest cosmological observations (Planck + BAO + SZ + WL + lensing + $H_0$) gives $N_{\rm eff}=3.30^{+0.12}_{-0.20}$ and $m_{\nu,\rm sterile}^{\rm eff}<0.242$~eV, indicating that the detection of $\Delta N_{\rm eff}>0$ is at the 1.27$\sigma$ level and the mass of sterile neutrino is very light (see \cite{deHolanda:2010am} for a study of very light sterile neutrinos). In this paper, we show that a search for sterile neutrinos in the HDE cosmology also favors a massless (or very light) sterile neutrino, with the detection of $\Delta N_{\rm eff}>0$ at the 2.75$\sigma$ level.

\begin{table}
\caption{\label{tab4}Fitting results for the HDE+$N_{\rm eff}$+$m^{\rm eff}_{\rm \nu,sterile}$ model from the constraint of the data combination CMB (Planck TT,TE,EE+$\tau$)+BAO+SN+RSD+WL+lensing+$H_{0}$. Note that the mass of sterile neutrino $m^{\rm eff}_{\rm \nu,sterile}$ is in units of eV and the Hubble constant $H_{0}$ is in units of $\rm km  ~s^{-1} ~Mpc^{-1}$.}
\centering
\begin{tabular}{ccc}

\hline\hline
$\Omega_{\rm b}h^2$&$0.02242^{+0.00014}_{-0.00017}$\\
$\Omega_{\rm c}h^2$&$0.1164^{+0.0045}_{-0.0042}$\\
$100\theta_{\rm MC}$&$1.04085^{+0.00036}_{-0.00031}$\\
$\tau$&$0.064\pm0.008$\\
$n_s$&$0.9676^{+0.0043}_{-0.0067}$\\
${\rm{ln}}(10^{10}A_s)$&$3.063\pm0.016$\\
$c$&$0.590^{+0.025}_{-0.030}$\\
$m_{\nu,{\rm{sterile}}}^{\rm{eff}}$&$<0.902$\\
$N_{\rm eff}$&$<3.35$\\
$\Omega_{\rm m}$&$0.2949^{+0.0076}_{-0.0082}$\\
$H_0$&$69.6\pm0.9$\\
$\sigma_8$&$0.788^{+0.026}_{-0.019}$\\

\hline\hline
\end{tabular}
\end{table}

\section{Discussion}\label{discussion}

\begin{table*}
\caption{\label{tab1} Fitting results for the HDE+$N_{\rm eff}+m_{\nu,\rm sterile}^{\rm eff}$ model from different data combinations. Here CMB refers to Planck TT+lowP. Note that $m_{\nu,{\rm{sterile}}}^{\rm{eff}}$ is in units of eV and $H_{0}$ is in units of $\rm km\;s^{-1} Mpc^{-1}$.}
\centering
\begin{tabular}{ccccccccc}
\hline
\hline
       Data&CMB &CMB+BAO&CMB+SN&CMB+RSD&CMB+WL&CMB+lensing&CMB+$H_{0}$\\
\hline

$\Omega_bh^2$&$0.02233^{+0.00027}_{-0.00023}$&$0.02279^{+0.00030}_{-0.00036}$&$0.02223^{+0.00021}_{-0.00026}$&$0.02263^{+0.00025}_{-0.00034}$&$0.02251\pm0.00025$&$0.02240^{+0.00027}_{-0.00029}$&$0.02238^{+0.00025}_{-0.00028}$\\
$\Omega_ch^2$&$0.1195^{+0.0078}_{-0.0034}$&$0.1225\pm0.0044$&$0.1154^{+0.0099}_{-0.0086}$&$0.1164^{+0.0081}_{-0.0058}$&$0.1190\pm0.0042$&$0.1218^{+0.0035}_{-0.0033}$&$0.1219^{+0.0042}_{-0.0036}$\\
$100\theta_{\rm MC}$&$1.04061^{+0.00051}_{-0.00056}$&$1.04071^{+0.00056}_{-0.00055}$&$1.04052^{+0.00049}_{-0.00056}$&$1.04096^{+0.00056}_{-0.00050}$&$1.04081\pm0.00050$&$1.04058^{+0.00050}_{-0.00049}$&$1.04051\pm0.00051$\\
$\tau$&$0.080^{+0.020}_{-0.022}$&$0.110\pm0.023$&$0.070^{+0.007}_{-0.006}$&$0.091^{+0.021}_{-0.025}$&$0.083^{+0.020}_{-0.032}$&$0.079^{+0.020}_{-0.022}$&$0.085\pm0.021$\\
$n_s$&$0.9679^{+0.0076}_{-0.0104}$&$0.9930\pm0.0140$&$0.9617^{+0.0064}_{-0.0080}$&$0.9820^{+0.0100}_{-0.0180}$&$0.9749^{+0.0083}_{-0.0108}$&$0.9742^{+0.0090}_{-0.0131}$&$0.9708^{+0.0085}_{-0.0130}$\\
${\rm{ln}}(10^{10}A_s)$&$3.101^{+0.025}_{-0.043}$&$3.161\pm0.048$&$3.079^{+0.011}_{-0.015}$&$3.115^{+0.042}_{-0.054}$&$3.103^{+0.041}_{-0.055}$&$3.098^{+0.039}_{-0.043}$&$3.113\pm0.042$\\
$c$&$0.454^{+0.051}_{-0.188}$&$0.640^{+0.069}_{-0.092}$&$0.634^{+0.044}_{-0.062}$&$0.610^{+0.071}_{-0.092}$&$0.418^{+0.038}_{-0.132}$&$0.529^{+0.079}_{-0.246}$&$0.489^{+0.043}_{-0.061}$\\
$m_{\nu,{\rm{sterile}}}^{\rm{eff}}$&$<1.305$&$<0.282$&$<0.137$&$<1.132$&$<0.765$&$<0.663$&$<0.839$\\
$N_{\rm eff}$&$<3.55$&$<4.09$&$<3.37$&$<3.87$&$<3.65$&$<3.67$&$<3.65$\\
$\Omega_m$&$0.240^{+0.037}_{-0.095}$&$0.281\pm0.013$&$0.338^{+0.017}_{-0.018}$&$0.283^{+0.016}_{-0.013}$&$0.210^{+0.025}_{-0.068}$&$0.266^{+0.054}_{-0.110}$&$0.272^{+0.015}_{-0.017}$\\
$H_0$&$80.7^{+15.6}_{-9.4}$&$72.4^{+1.9}_{-2.2}$&$65.9^{+1.4}_{-1.5}$&$71.4^{+1.6}_{-2.2}$&$85.0^{+15.0}_{-4.5}$&$77.1^{+10.4}_{-14.9}$&$73.9\pm1.9$\\
$\sigma_8$&$0.886^{+0.099}_{-0.096}$&$0.836\pm0.028$&$0.749^{+0.028}_{-0.041}$&$0.791^{+0.032}_{-0.024}$&$0.912^{+0.102}_{-0.070}$&$0.856^{+0.092}_{-0.105}$&$0.844^{+0.040}_{-0.031}$\\

\hline
\hline
\end{tabular}\end{table*}

In this section, we make some further discussions for the fitting results derived in the last section. 

From Table~\ref{tab2}, for the constraints without the $H_0$ prior, we notice that the reionization optical depth $\tau$ will become larger when the sterile neutrinos are considered in the $\Lambda$CDM model (i.e., $\tau=0.063\pm 0.013$ for $\Lambda$CDM and $\tau=0.081^{+0.017}_{-0.019}$ for $\Lambda$CDM+$N_{\rm eff}$+$m^{\rm eff}_{\rm \nu,sterile}$); and, $\tau$ will also become larger when the HDE model is considered (i.e., $\tau=0.097\pm 0.016$). For the HDE+$N_{\rm eff}$+$m^{\rm eff}_{\rm \nu,sterile}$ model, we obtain $\tau=0.114\pm 0.019$ for this case (see Table~\ref{tab2}), and it becomes even larger when the $H_0$ prior is added in the fit, $\tau=0.119^{+0.019}_{-0.020}$ (see Table~\ref{tab3}). Such a high $\tau$ value is evidently in tension with the new measured $\tau$ value, $0.055\pm0.009$, as found by the High Frequency Instrument (HFI) of the Planck satellite \cite{Aghanim:2016yuo} (in this case, the $H_0$ tension is at more than 3$\sigma$). On the other hand, we did not use the Planck polarization data at high multipoles in the cosmological fits of the last section. Here we also wish to see what will happen when the Planck polarization data are added in the fit. 

Now we use the Planck temperature and polarization data to do the analysis, where the Planck temperature and polarization data at low multipoles are replaced with a Gaussian prior on the reionization optical depth $\tau=0.055\pm0.009$, and we denote this dataset as ``Planck TT,TE,EE+$\tau$''. We combine this CMB dataset with the other datasets (BAO+SN+RSD+WL+lensing+$H_0$), to constrain the HDE+$N_{\rm eff}$+$m^{\rm eff}_{\rm \nu,sterile}$ model, with the constraint results shown in Table~\ref{tab4}. We find that, in this case, the $\tau$ value becomes evidently smaller, i.e., we obtain $\tau=0.064\pm0.008$. For the HDE model parameter, we have $c=0.590^{+0.025}_{-0.030}$ in this case (in the case without polarization data and the $\tau$ prior, we have $c=0.728^{+0.052}_{-0.062}$; see Table~\ref{tab3}). So we see that when the polarization data and $\tau$ prior are considered, the constraint precision for the cosmological parameters will be largely enhanced. For the sterile neutrino parameters, we have $N_{\rm eff}<3.35$ and $m^{\rm eff}_{\rm \nu,sterile}<0.902$ eV. Thus, in this case, $N_{\rm eff}$ cannot be tightly constrained, and we can only obtain an upper limit for it; the upper limit on $m^{\rm eff}_{\rm \nu,sterile}$ becomes much larger. Since the fit value of $N_{\rm eff}$ changes greatly (it only has upper limit and becomes much smaller), the spectral index $n_s$ becomes smaller, $n_s=0.9676^{+0.0043}_{-0.0067}$, due to the positive correlation between $n_s$ and $N_{\rm eff}$ (for the case without polarization data and $\tau$ prior, $n_s=1.004\pm 0.011$; so this time, the spectrum becomes red again). In this case, the $H_0$ value becomes smaller, i.e., $H_0=69.6\pm 0.9$ km~s$^{-1}$~Mpc$^{-1}$, in tension with R16 at 1.73$\sigma$, so actually the tension becomes more severe (in the case of Table~\ref{tab3}, the $H_0$ tension is only at 0.85$\sigma$).

Next, we discuss the issue of the small-scale matter fluctuation amplitude in the HDE+$N_{\rm eff}$+$m^{\rm eff}_{\rm \nu,sterile}$ model. Recently, the first tomographic cosmic shear analysis of the Kilo Degree Survey (KiDS) using almost one third of the final data volume ($\sim 450$ deg$^2$) was presented \cite{Hildebrandt:2016iqg}. They found a best-fit value for $S_8\equiv \sigma_{8}\sqrt{\Omega_{m}/0.3}=0.745\pm 0.039$ assuming a flat $\Lambda$CDM model using weak external priors \cite{Hildebrandt:2016iqg}. This result is in tension with the Planck 2015 result at the 2.3$\sigma$ level but consistent with previous cosmic shear analyses and a number of other literature measurements. For the HDE+$N_{\rm eff}$+$m^{\rm eff}_{\rm \nu,sterile}$ model, using the CMB+BAO+SN+RSD+WL+lensing+$H_0$ data combination, we obtain $\sigma_{8}=0.814\pm0.013$ (see Table~\ref{tab3}); we also calculate the corresponding $S_8$ value and we obtain $S_{8}=0.802^{+0.019}_{-0.014}$. Though the KiDS-450 result $S_8=0.745\pm 0.039$ is a fit value for the $\Lambda$CDM model, it is derived from a new measurement of cosmic shear and consistent with the previous measurements of cosmic shear; thus we compare our result with the KiDS-450 result. We find a tension at the 1.9$\sigma$ level between the two. As discussed in the Planck 2015 paper \cite{Ade:2015xua}, for solving the tension with weak lensing measurements, the scheme with sterile neutrinos offers  only a marginal improvement compared to the base $\Lambda$CDM model. Here we have made an analysis for the HDE cosmology with sterile neutrinos, and we find that the same conclusion remains. 

Our calculations are mainly based on the Planck observations. The combination of the Planck data with other astrophysical observations is very important because the Planck data alone cannot provide tight constraints for most cases. In this paper, we considered several important external astrophysical observations and we combined these datasets together to make the analyses. Now we wish to provide a complementary calculation for the HDE+$N_{\rm eff}$+$m^{\rm eff}_{\rm \nu,sterile}$ model, in which we combine the Planck data (Planck TT+lowP) with the single external dataset separately (i.e., adding a single dataset per time). The results are shown in Table~\ref{tab1}. In this table, the results are listed for the cases of CMB, CMB+BAO, CMB+SN, CMB+RSD, CMB+WL, CMB+lensing, and CMB+$H_0$. Here CMB refers to the Planck TT+lowP data. From this calculation, we can see that the external astrophysical observations are basically consistent with the Planck data. The Planck data alone cannot provide tight constraints and some of the external datasets can play an important role in breaking the degeneracies of Planck data, but combining the Planck data with any single one dataset still cannot provide tight enough constraints for cosmological parameters. Thus, the all-data analysis is necessary for our task. Of course, however, one should be aware that there are still some residual systematics in these external astrophysical observations, which will lead to some potential biases in the global fit results. Currently we cannot completely avoid the biases in our global fit analysis.

\section{Conclusion}\label{sec:conclu}

We consider the massive sterile neutrinos in holographic dark energy cosmology in this paper. We have two main aims: (i) We wish to investigate if the tension between the latest local measurement of the Hubble constant and the Planck observation can be effectively relieved by considering both sterile neutrinos and holographic dark energy. (ii) We wish to search for sterile neutrinos in the holographic dark energy cosmology with the latest cosmological observations. To perform the analysis, we employ the current observations, including the Planck CMB temperature power spectrum, the BAO measurements, the SN data, the RSD measurements, the shear data of WL observation, the Planck lensing measurement, and the latest direct measurement of $H_0$ as well.

We show that, compared to the $\Lambda$CDM cosmology, the HDE cosmology with sterile neutrinos can relieve the tension much better. Using the CMB+BAO+SN+RSD+WL+lensing data combination to constrain the HDE+$N_{\rm eff}$+$m_{\nu,\rm sterile}^{\rm eff}$ model, we obtain $N_{\rm eff}<4.03$ and $m_{\nu,\rm sterile}^{\rm eff}<0.225\,\rm eV$. In this case, we find that the tension of $H_0$ is relieved to be at the only 1.43$\sigma$ level. This indicates that considering both sterile neutrinos and holographic dark energy can indeed largely relieve the tension. But in this case (without the inclusion of the $H_0$ measurement in the analysis), only the upper limits of $N_{\rm eff}$ and $m_{\nu,\rm sterile}^{\rm eff}$ can be given.

We further include the $H_0$ measurement in the global fit of the HDE+$N_{\rm eff}$+$m_{\nu,\rm sterile}^{\rm eff}$ model, and we find that a hint of the existence of sterile neutrinos in the HDE cosmology can be given. Under the constraint of the all-data combination, we obtain $N_{\rm eff}= 3.76\pm0.26$ and $m_{\nu,\rm sterile}^{\rm eff}< 0.215\,\rm eV$, indicating that the detection of $\Delta N_{\rm eff}>0$ in the HDE cosmology is at the $2.75\sigma$ level and the massless or very light sterile neutrino is favored by the current data.

\acknowledgments
This work was supported by the National Natural Science Foundation of China (Grants No.~11522540 and No.~11690021), the National Program for Support of Top-Notch Young Professionals, and the Provincial Department of Education of Liaoning (Grant No.~L2012087).

\end{document}